\documentclass{article}
\usepackage{spconf,amsmath,graphicx}
\usepackage{hyperref}
\usepackage{caption,subcaption}
\usepackage{amsfonts, cleveref}
\usepackage{booktabs}
\usepackage[bottom]{footmisc}
\usepackage[inline]{enumitem}

\def\X{{\boldsymbol X}}

\def\Z{{\boldsymbol Z}}
\def\H{{\boldsymbol H}}
\def\C{{\boldsymbol C}}
\def\S{{\boldsymbol S}}
\def\R{{\mathbb R}}
\def\L{{\cal L}}

\title{AGAIN-VC: A One-shot Voice Conversion using \\
Activation Guidance and Adaptive Instance Normalization}
\name{Yen-Hao Chen, Da-Yi Wu, Tsung-Han Wu, Hung-yi Lee}
\address{College of Electrical Engineering and Computer Science, National Taiwan University, Taiwan}
\email{\{r07921112, r07922119, r07942145, hungyilee\}@ntu.edu.tw}
\begin{document}
\ninept

\maketitle

\begin{abstract}
    Recently, voice conversion (VC) has been widely studied. Many VC systems use disentangle-based learning techniques to separate the speaker and the linguistic content information from a speech signal. Subsequently, they convert the voice by changing the speaker information to that of the target speaker. To prevent the speaker information from leaking into the content embeddings, previous works either reduce the dimension or quantize the content embedding as a strong information bottleneck. These mechanisms somehow hurt the synthesis quality. In this work, we propose \textbf{AGAIN-VC}, an innovative VC system using \textbf{A}ctivation \textbf{G}uidance and \textbf{A}daptive \textbf{I}nstance \textbf{N}ormalization. AGAIN-VC is an auto-encoder-based model, comprising of a single encoder and a decoder. With a proper activation as an information bottleneck on content embeddings, the trade-off between the synthesis quality and the speaker similarity of the converted speech is improved drastically. This one-shot VC system obtains the best performance regardless of the subjective or objective evaluations.
\end{abstract}
\section{Introduction}
\label{sec:intro}
Voice conversion (VC) is to convert the voice of a source speech to that of a target speech yet meanwhile remains the linguistic content. 
Previous works use different techniques, including generative adversarial networks (GAN),  flow, and variational auto-encoder (VAE), etc., to tackle many-to-many voice conversion \cite{kaneko2018cyclegan, kaneko2019cyclegan, kameoka2018stargan, kaneko2019stargan, patel2019novel, serra2019blow, hsu2016voice, hsu2017voice, huang2018voice, kameoka2019acvae, kaneko2017parallel, Chou2018, goodfellow2014generative, kingma2013auto}. 

Recently, there are several works perform one-shot VC, which means that the source and the target speakers can be both unseen in the inference phase \cite{qian2019autovc, qian2020f0, qian2020unsupervised, Chou2019, wu2020one, wu2020vqvc+}. AutoVC pre-trains a speaker encoder using GE2E loss and designs a dimension bottleneck on the content encoder; the tuned bottleneck successfully separates the speaker and the content \cite{qian2019autovc, wan2018generalized}. AdaIN-VC utilizes a variational auto-encoder consisting of a speaker encoder, a content encoder, and a decoder \cite{Chou2019}. With adaptive instance normalization (AdaIN), speaker information and content information can be separated well \cite{huang2017arbitrary}. VQVC+ regards linguistic content information as discrete representations and uses vector quantization (VQ) to extract content embeddings \cite{wu2020vqvc+, van2017neural}. Besides, VQVC+ applies U-net as the network architecture for  better reconstruction \cite{ronneberger2015u}.

Those methods mentioned above achieve one-shot VC, yet one-shot VC is still a challenging task. AutoVC has an obvious limitation that the pre-trained speaker encoder is merely trained for speaker verification. Hence, the robustness of generating speech is questionable. AdaIN-VC uses two independent encoders to extract speaker embeddings and content embeddings, respectively.
Nevertheless, we believe that the speaker encoder here is somewhat redundant; that is, their tasks could be done by a single encoder. 
Furthermore, although VQVC+ has a good performance on speaker conversion, content information is seriously damaged due to the discrete nature of VQ. 

In this work, we propose a novel VC system, \textbf{AGAIN-VC}, which stands for \textbf{A}ctivation \textbf{G}uidance and \textbf{A}daptive \textbf{I}nstance \textbf{N}ormalization. Our method improves the trade-off between the synthesis quality and the disentangling ability dramatically.

The main contributions of this work are listed below:
\begin{enumerate}
     \item Applying a single encoder for disentangling speaker and content information to improve the synthesis quality and reduce the model size simultaneously.
     \item Introducing activation guidance, applying an additional activation function as an information bottleneck to guide the content embeddings in continuous space, to improve the trade-off between the synthesis quality and the disentangling ability.
\end{enumerate}

\begin{figure}[ht]

\begin{minipage}[b]{1.0\linewidth}
  \centering
  \centerline{\includegraphics[width=7cm]{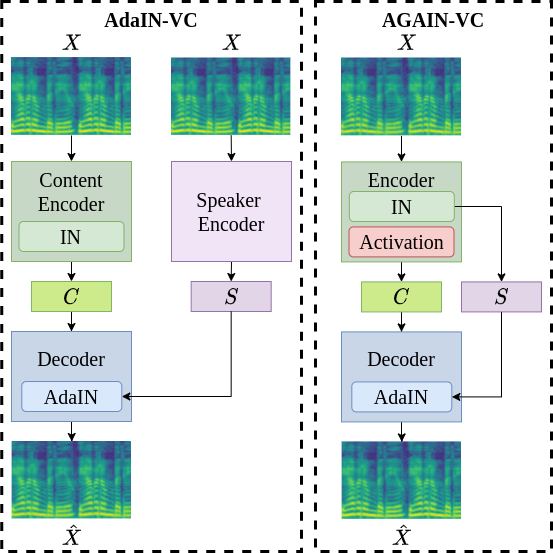}}
\end{minipage}
\caption{AdaIN-VC and AGAIN-VC. AdaIN-VC uses a content encoder and a speaker encoder, while AGAIN-VC uses only one encoder and an activation to guide the training.}
\label{fig:diff-adain}
\end{figure}

\begin{figure}[ht]

\begin{minipage}[b]{1.0\linewidth}
  \centering
  \centerline{\includegraphics[width=7cm]{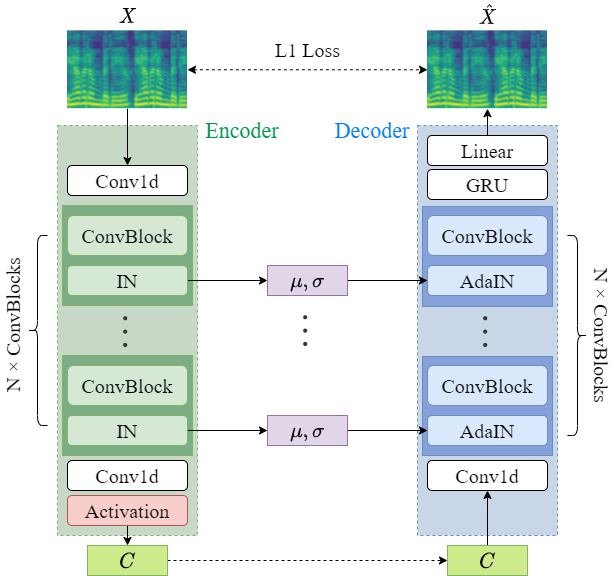}}
\end{minipage}
\caption{
AGAIN-VC architecture. The left part is the encoder, while the right part is the decoder. Note that L1 Loss is to make the input $\X$ and the output $\hat \X$ as close as possible.}
\label{fig:model}
\end{figure}


%

\section{Methods}
\label{sec:mehtods}

\subsection{System overview}
\label{ssec:autoencoder}
At the very beginning, we denote $\X \in  \R^{K\times T}$ as the input mel-spectrogram of an audio segment, where $K$ is the number of the frequency bins of the acoustic feature per frame, and $T$ represents the duration of the segment.

\Cref{fig:diff-adain} shows the model overview of AdaIN-VC and AGAIN-VC \cite{Chou2019}. AdaIN-VC treats the speaker information as the global style of a speech, which is supposed to be time-invariant.
On the contrary, the content information is regarded as the local one because of its time-varying characteristics.
In this point of view, instance normalization (IN) in AdaIN-VC is applied for the purpose of feature disentanglement \cite{huang2017arbitrary}. 
Compared with AdaIN-VC, AGAIN-VC uses only a single encoder to extract the speaker and the content representations. With this well-designed model architecture, a single encoder has better performance on disentangling the speaker and the content information. More specifically, instead of building one extra speaker encoder, we directly reuse the channel-wise mean $\mu$ and the channel-wise standard deviation $\sigma$ computed by IN layers as speaker embeddings.
Moreover, we introduce activation guidance (AG) to boost the VC performance, which will be explained in detail in \cref{ssec:act}. Last but not least, 
AGAIN-VC uses only a self-reconstruction loss  \begin{align}
     \L = ||\X-\hat \X||_1^1, \label{eq:loss}
 \end{align}
 where $\hat \X$ is the output of the auto-encoder.

\subsection{Style transfer using AdaIN}
\label{ssec:adain}
Given any input $\Z$, where $\Z \in \R^{K \times T}$, IN first computes the channel-wise mean $\mu$ and the channel-wise standard deviation $\sigma$. Then, the normalized representation can be obtained by
\begin{align}
    \text{IN}(\Z) = \frac{\Z - \mu(\Z)}{\sigma(\Z)}.
\end{align}
Note that the $\mu$ and $\sigma$ are time-invariant, and consequently, they are regarded as kinds of the global (speaker) representations. As shown in \cref{fig:model}, we add an IN layer in each encoder block to detach the global information from the hidden encoded representation. It is worth mentioning that the detached feature $\mu$ and $\sigma$ are reused in the adaptive instance normalization (AdaIN) layer during the decoding phase.
Suppose that the source and the target representations are $\H$ and $\Z$, respectively. The style transfer can be reached by two steps:
\begin{enumerate*}
  \item Extract the global features $\mu$ and $\sigma$ from $\Z$.
  \item Pass the $\H$, $\mu$, and $\sigma$ into an AdaIN layer, where AdaIN is defined as
\end{enumerate*}
\begin{align}
    \text{AdaIN}(\H, \mu(\Z), \sigma(\Z)) = \sigma(\Z) \text{IN}(\H)+\mu(\Z).
\end{align}
This process transfers the style of $\H$ to the style of $\Z$, while remains the content of $\H$ \cite{huang2017arbitrary,Chou2019}. 
As shown in \cref{fig:model}, we apply U-net architecture \cite{ronneberger2015u}. The input mel-spectrogram $\X$ is passed through several IN layers to eliminate its global (speaker) information. 
Meanwhile, the skip-connection structure passes the speaker embeddings $\mu$ and $\sigma$ of each layer
to the corresponding AdaIN layer in the decoder block for style transfer. Eventually, the generated mel-spectrogram $\hat \X$ is used to compute the reconstruction loss \eqref{eq:loss}.

\subsection{Activation guidance (AG)}
\label{ssec:act}
Instead of dimension reduction or vector quantization, we add an activation function as an information bottleneck to prevent the content embedding $\C$ from leaking speaker information. 
With an extra activation function, the range of content embeddings is somewhat restricted. While some activation functions (e.g. ReLU) are too harsh and thus ruin the important information encoded in embeddings, others (e.g. Sigmoid) moderately constraint the values without hurting the performance of reconstruction, acting as an outstanding information bottleneck. The detailed comparison between different activation functions is in \cref{sssec:comparison_act}.
With this additional bottleneck, models have to grasp more information from the speaker embeddings, which simply comprises of $\mu$ and $\sigma$.
However, $\mu$ and $\sigma$ are time-invariant; hence they have no ability to represent the content varying over time, meaning that the content information can never be carried by $\mu$ and $\sigma$.
In such circumstances, the global (speaker) information is much easier to learn for this model.
To some degree, this force content embeddings to guide themselves to learn more content information. In this way, speaker information cannot and should not be conveyed by content embeddings anymore; content information is forced to be carried by content embeddings only concurrently. In short, our proposed method guides the flow of information in the right direction. 

\section{Implementation Details}
\label{sec:illust}
\subsection{Dataset and data preprocessing}
\label{ssec:dset}
There are 109 English speakers in VCTK, which is the training corpus in this work \cite{yamagishi2019cstr}. We randomly selected 80 speakers for training, and the remaining speakers are for evaluation. Also, for each speaker, there were 200 utterances chosen arbitrarily during training.
As for the data preprocessing, the waveforms were converted to 22050Hz, and we trimmed the silence at the head and tail of each audio clip subsequently.  Thereafter, the audio clips were converted to mel-spectrogram with 1024 STFT window size, 256 hop length, and 80 mel bins according to the configuration in MelGAN \cite{kumar2019melgan}. Additionally, we set the segment length of each mel-spectrogram to 128, which is 1.5 seconds or so.

\subsection{Training details}
The proposed model was trained using ADAM optimizer, initialized with a 0.0005 learning rate, $\beta_1=0.9$, and $\beta_2=0.999$ \cite{kingma2014adam}. The batch size was set to 32, and the number of training steps was 100k.
Refer to the implementation code for more details.
\footnote{\href{https://github.com/KimythAnly/AGAIN-VC}{https://github.com/KimythAnly/AGAIN-VC}}.
\subsection{Vocoder}
\label{ssec:vocoder}
Because the output of our proposed model is mel-spectrogram, we need a vocoder to convert the mel-spectrogram to the waveform.
Considering the quality and the inference speed of waveform generation, we utilize a pre-trained MelGAN as the vocoder \cite{kumar2019melgan}.

\section{Experiments}
\label{sec:exp}

\subsection{Evaluation on activation guidance}
\label{ssec:eval-act}
To examine how much speaker information is in the content embedding $\C$, a speaker classifier was trained on $\C$ to probe the speaker identity for each model.
This speaker classifier is made up of three Conv1d layers and three ReLU layers alternatively, followed by a linear layer.
\subsubsection{Effect of activation functions}

\begin{figure}[t]

\begin{minipage}[b]{1.0\linewidth}
  \centering
  \centerline{\includegraphics[width=7cm]{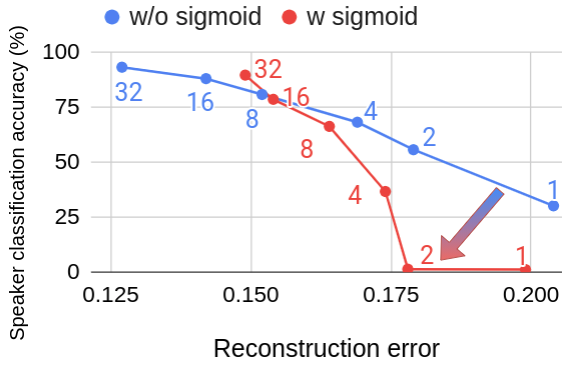}}
\end{minipage}
\caption{The trade-off between the reconstruction error and the speaker classification accuracy on $\C$. The numbers here represent the channel size of $\C$. Besides, the arrow stands for the direction of improvement of the models we do care.}
\label{fig:trade-off}
\end{figure}

In general, there is a trade-off between the reconstruction error (speech quality) and the speaker classification accuracy (disentangling ability); both of them should be as low as possible.
In \cref{fig:trade-off}, the reconstruction error becomes large, and the speaker classification accuracy becomes low as the channel size decreases.
Moreover, with a sigmoid function as AG, most points move down and slightly right.

From a macro perspective, the trade-off curve is improved from blue to red; speaker classification accuracy of all models decreases drastically while the reconstruction error of most models increases subtly. Also, what we really care about is those models with the speaker classification accuracy less than 70\% because empirically, models with accuracy higher than 70\% usually fail to convert the audio clips. It is evident that these kinds of models with sigmoid function perform much better.

\subsubsection{Different activation functions}
\label{sssec:comparison_act}
In addition to sigmoid function, we also explore several activation functions, such as ReLU, ELU, Tanh, and several variant sigmoid functions:
\begin{align}
    \text{Sigmoid}(x) = \frac{1}{1+\exp(- \alpha x)},
\end{align}
where $\alpha$ is a hyperparameter. Besides the reconstruction error and the speaker classification accuracy on $\C$, we used an extra speaker classifier to probe the speaker identity on speaker embeddings $\S$, which simply comprises of $\mu$ and $\sigma$. Ideally, the accuracy should be as high as possible on $\S$, while it should be as close to 1.25\% as possible on $\C$ since there are 80 speakers in the dataset. The result is shown in \cref{tab:acts}.

First of all, we found that the speaker classification accuracy on $\S$ among activation functions are all beyond 90\% yet with a slight difference; it seems that this metric exactly has nothing to do with the trade-off between the speaker classification accuracy on $\C$ and the reconstruction error.
On the contrary, it is obvious that not all activation functions benefit the training; ELU does hurt the performance to some extent. Likewise, ReLU hinders the reconstruction more severely yet with a bit better probing accuracy on $\C$.
In particular, half of the input of the activation functions are negative, and these two activation functions ruin the information encoded in the negative values. Thus, both of them have higher reconstruction error.
Yet the slightly lower speaker classification accuracy of ReLU on $\C$ might be due to the information loss of those negative values. In the second place, Sigmoid ($\alpha=1$) obtains much better accuracy on $\C$ while achieving a similar reconstruction error. As a result, we choose several $\alpha$ for further analysis. The results imply that $\alpha=0.1$ may be the best hyperparameter, being able to filter out the speaker information from content embeddings and surprisingly yielding the lowest reconstruction error simultaneously. Specially note that Tanh is de facto the shifted Sigmoid ($\alpha=2$). 

To conclude, Sigmoid with well-chosen $\alpha$ is suitable for VC since it is essentially a mild restriction, whereas it does not ruin the encoded information like other activation functions.

\captionsetup{belowskip=12pt,aboveskip=4pt}
\begin{table}[t]
    \small
    \center
    \caption{Comparison between the models with different activation functions. $\C$ and $\S$ are the speaker classification accuracy on content embeddings and speaker embeddings, respectively, and Rec. represents the reconstruction error. * is our proposed method.}
    \begin{tabular}{ l| c c c c} 
    \toprule
    Activation & $\C$ (Acc.\%) $\downarrow$ & $\S$ (Acc.\%) $\uparrow$ & Rec. $\downarrow$ \\
    \midrule
    None & 68.6 & 92.6 & 0.161 \\ 
    ReLU \ & 51.9 & 92.2 & 0.174 \\ 
    ELU \ & 69.2 & 91.5 & 0.167 \\ 
    Tanh \ & 57.3 & 91.7 & 0.165 \\ 
    Sigmoid ($\alpha=1$) & 30.5 & 90.0 & 0.167 \\ 
    Sigmoid ($\alpha=0.1$) $^*$ & \textbf{1.7} & \textbf{93.2} & \textbf{0.151} \\
    Sigmoid ($\alpha=0.01$) & \textbf{1.7} & 91.1 & 0.222 \\
    \bottomrule
    \end{tabular}
    \label{tab:acts}
\end{table}

\subsection{Single encoder}
\label{ssec:1enc}
To verify our claim that using two encoders is redundant and unnecessary, we also trained AGAIN-VC with ``two independent encoders'', which is reminiscent to AdaIN-VC. For these models (2-Enc), the content encoder and the speaker encoder use the same architecture while their weights are independent. Note that the 2-Enc models were trained with or without activation guidance respectively. The result is shown in \cref{tab:1enc}. The size of the models with a single encoder (1-Enc) is about 30\% less than those with two encoders (2-Enc); nonetheless, the performance are almost identical. Most importantly, 1-Enc-sig, the proposed model, obtains the best performance under all evaluation metrics.
\captionsetup{belowskip=12pt,aboveskip=4pt}
\begin{table}[t]
    \center
    \small
    \caption{Comparison between the models using a single encoder (1-Enc) and those with two encoders (2-Enc). Note that $\C$ and $\S$ represent the speaker classification accuracy on content embeddings and speaker embeddings, respectively; Rec. is the reconstruction error, and the last column is the model size. Also, ``-sig'' represents that sigmoid ($\alpha=0.1$) is added on $\C$; * is our proposed method.}
    \begin{tabular}{ l| c c c c} 
    \toprule
     & $\C$ (Acc.\%) $\downarrow$ & $\S$ (Acc.\%) $\uparrow$ &  Rec. $\downarrow$ & Size $\downarrow$ \\
    \midrule
    1-Enc & 68.6 & 92.6 & 0.161 & \textbf{9.5 M }\\ 
    2-Enc & 67.2 & 91.5 & 0.167 & 13.5 M \\ 
    1-Enc-sig $^*$ & \textbf{1.7} & \textbf{93.2} & \textbf{0.151} & \textbf{9.5 M} \\ 
    2-Enc-sig & 1.8 & 92.7 & 0.154 & 13.5 M \\ 
    \bottomrule
    \end{tabular}
    \label{tab:1enc}
\end{table}
\subsection{Subjective evaluations}
\label{ssec:subjective}

\begin{figure}[t]
\begin{subfigure}[b]{0.48\linewidth}
  \centering
  \centerline{\includegraphics[width=3.5cm]{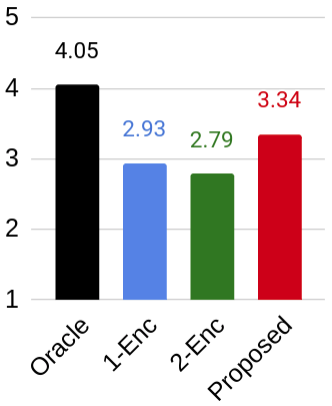}}
  \caption{Naturalness test (MOS)}
  \label{fig:mos}
\end{subfigure}\hfill
\begin{subfigure}[b]{0.48\linewidth}
  \centering
  \centerline{\includegraphics[width=5cm]{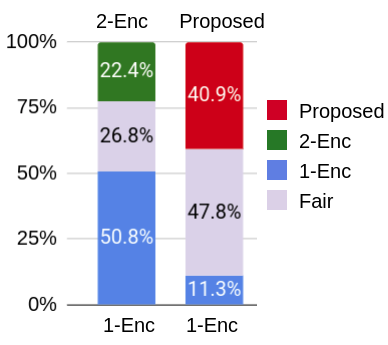}}
  \caption{Pairwise similarity test}
  \label{fig:sim}
\end{subfigure}\hfill
\caption{Subjective evaluations}
\label{fig:subjective}
\end{figure}

We performed two subjective evaluations, naturalness test and pairwise similarity test, on one-shot voice conversion for three models, 1-Enc, 2-Enc, and Proposed. The results are shown in \cref{fig:mos} and \cref{fig:sim}, respectively.
We measured the naturalness with the help of mean opinion score (MOS). MOS ranges from 1 to 5, and is the higher the better. Here we regarded the speech converted from real mel-spectrogram using MelGAN vocoder as the oracle.
As for the pairwise similarity test, a reference (target) speech and two converted speech generated by two different models were given. The participants were asked to select the converted speech whose speaker is more similar to the reference's.
The results in \cref{fig:subjective} indicate that 1-Enc outperforms 2-Enc regardless of naturalness or similarity. It verifies that using a single encoder is absolutely sufficient and more effective for VC. Furthermore, the proposed method is better than 1-Enc, implying that applying a proper activation function to content embeddings as an information bottleneck does benefit models. More synthesized examples can be found in our demo page \footnote{
\href{https://kimythanly.github.io/AGAIN-VC-demo/index}{https://kimythanly.github.io/AGAIN-VC-demo/index}
}.

\section{Conclusion}
In this work, we proposed AGAIN-VC, a one-shot VC with a single encoder and the proposed activation guidance (AG) technique.
The experimental results showed that a single encoder is sufficient for feature disentanglement, whereas AG is an extraordinary information bottleneck preventing information leaking.
Above all, human evaluation highly supports the effectiveness of utilizing a single encoder and activation guidance. 
\label{sec:conclusion}

\small
\bibliographystyle{IEEEbib}
\bibliography{strings,refs}

\end{document}